\documentclass[
    ,final            
  ]
  {aipproc}

\layoutstyle{6x9}

\newcommand{\prd}{PRD}
\newcommand{\Ms}{M_{\odot}}
\newcommand{\apjl}{ApJL}

\begin{document}

\title{Automatic Bayesian inference for LISA data analysis strategies}


\author{Alexander Stroeer}{
  address={School of Physics and Astronomy,
University of Birmingham
Edgbaston,
Birmingham,
B15 2TT,
United Kingdom}
}
\author{Jonathan Gair}{
  address={Institute of Astronomy,
Madingley Road,
Cambridge,
CB3 0HA, United Kingdom}
}
\author{Alberto Vecchio}{
  address={School of Physics and Astronomy,
University of Birmingham
Edgbaston,
Birmingham,
B15 2TT,
United Kingdom}
}

\begin{abstract}
We demonstrate the use of automatic Bayesian inference for the analysis of LISA data sets. In particular we describe a new automatic Reversible Jump Markov Chain Monte Carlo method to evaluate the posterior probability density functions of the {\em a priori} unknown number of parameters that describe the gravitational wave signals present in the data. We apply the algorithm to a simulated LISA data set containing overlapping signals from white dwarf binary systems and to a separate data set containing a signal from an extreme mass ratio inspiral. We demonstrate that the approach works well in both cases and can be regarded as a viable approach to tackle LISA data analysis challenges.
\end{abstract}

\maketitle


\section{Introduction}

The Laser Interferometer Space Antenna (LISA) is a proposed all-sky gravitational wave observatory with an expected launch date of 2015~\cite{lisa_ppa}. It will consist of a 3-arm laser interferometer in heliocentric orbit trailing the Earth by 20 degrees and counter-rotating with a period of 1 yr in a plane inclined by 60 degrees with respect to the ecliptic. The instrument will be sensitive to gravitational waves (GWs) in the frequency band 0.1 mHz - 0.1 Hz and is expected to observe a great variety of sources, ranging from known galactic binary systems to high redshift binaries harbouring a massive black hole~\cite{Cutler2002a}.

One of the challenges prior to the launch of the mission is to develop data analysis strategies able to provide accurate estimates of the parameters of the sources that generate the gravitational waves present in the data stream. In fact the LISA data set will contain a large ({\em a priori} unknown) number of partially overlapping signals. Here we consider an approach to the analysis within the framework of Bayes' inference~\cite{Brett1988a}. We use techniques known as Reversible Jump Markov Chain Monte Carlo (RJMCMC) methods~\cite{Green2003a} to evaluate the posterior probability density function (PDF) of the source parameters when the date set contains an unknown number of signals drawn from a large parameter space. Markov Chain Monte Carlo Methods have already been considered in several LISA data analysis case studies, {\em e.g.}~\cite{cornish05,Um2005b,cornish06b}. In these proceedings we report the development and application of an ``automatic'' RJMCMC sampler ({\em e.g.}~\cite{Green2003a,Hast2004a}), in which the user just provides a model for the signal(s) expected in the data set and the algorithm automatically returns the relevant posterior PDFs.

``Automatism'' is a challenge for the theory behind RJMCMC sampling. We implement Metropolis-Hastings sampling, with the underlying proposal density for the Markov Chain state transition as the object of an automatic optimisation according to the Adaptive Acceptance Probability (AAP) technique. AAP tries to find the optimal proposal density to achieve a target acceptance rate of 0.25 for Markov Chain state transitions: in this way a quarter of all proposed transitions should be accepted for a new Markov Chain state. AAP suggests bolder transitions if the estimated acceptance rate is larger than 0.25 or stricter proposals if it is smaller than 0.25. This method has proven to be simple in its implementation and computationally effective in a number of applications. In the case of LISA, the number of signals in the data set (and therefore the number of parameters, {\em i.e., dimensions}, that characterise the problem at hand) is unknown. We implement AAP for fixed dimension updates (jumps within a proposed model)~\cite{atch2003a}, but also for trans-dimensional updates (jumps between proposed models)~\cite{Hast2004a}. In addition we support AAP optimisation by finite mixture fits to the joint posterior density function. In order to increase efficiency, the sampler is implemented in parallel on loosely connected CPUs (such as Beowulf-type clusters). This is achieved by running independent Markov chain samples in parallel, which are then combined after the run, as discussed in~\cite{rosenthal-parallel}. A more detailed discussion of the implementations and techniques, including some expanded examples, will be reported elsewhere.

Here we describe the application of this analysis approach to two case studies: (i) the analysis of a data set containing two highly overlapping monochromatic DWD signals (actual number of signals and noise level in the data set unknown prior to the analysis); (ii) the analysis of a data set containing a signal from an extreme mass ratio inspiral (EMRI) (number of signals known but noise level unknown in the data set prior to the analysis). We demonstrate that the sampler is able to process these two data sets automatically to return posterior density functions that are consistent with the actual parameter values. We discuss the performance of this approach and future work that is necessary to apply this strategy to a realistic LISA data stream. 

\section{Overlapping signals from double white dwarf binaries}

The data $D$ recorded by the LISA observatory will consist of a superposition $H^{(K)}$ of $K$ signals $h^{(k)}$ (with $k = 1,\dots, K$) plus noise $n$, discretely sampled in time ($D_j=H_j^{(K)} + n_j$). The signals $h^{(k)}$ in our first case study are generated by double white dwarfs (DWD) (see e.g. \cite{Stroeer2005a}). For simplicity we assume that the GWs are exactly monochromatic in the source reference frame; each of the $h^{(k)}$ is described by the 7-dimension parameter vector $\vec{\theta}=\left\{ A_c, \Phi_0, f, \vartheta_N, \varphi_N, \vartheta_L, \varphi_L \right\}$, consisting of the amplitude, the phase at the beginning of the observation, the frequency of the incoming GW and the four angles that describe the source position and orientation in polar coordinates. For simplicity, we model the LISA detector output using the 2 Michelson observables (hI, hII) described by Cutler \cite{Cutler1998a}. We model the noise as Gaussian, stationary and white, described by the parameter $\sigma_n$, the standard deviation of the random process (the mean is set to zero).

In this analysis, the number of signals, $K$, is unknown as well as the noise level $\sigma_n$.  The target of RJMCMC sampling is to evaluate the joint posterior $p(K,\vec{\theta}^{(k)} \{k = 1,\dots,K\}, \sigma_n|D)$, see \cite{Green2003a} for details, and any relevant marginalised posterior density function (PDF) that one wishes to compute. The marginalised PDFs are the ones that we quote in the results. We created a data set for an observation time of half a year, with a sampling time of 100 seconds. We injected two DWD signals in the data streams, and searched for up to three DWD signals ($K_{max} = 3$). The maximum dimension for the parameter space is therefore 22 ($= 3\times 7 + 1$). The first signal was chosen to have parameters similar to the verification binary ``ES Cet'' \cite{Stroeer2006e}; the second signal was chosen to overlap with a match of 0.77 to mimic a ``confusion source''. The frequency of the signals differed by only half a frequency resolution bin and the two binaries had the same location and orientation in the sky. The signal-to-noise ratio (SNR) for the confusion source was a factor of three smaller than the other source. We sampled $10^7$ Markov chain states with a sampler designed  to distribute these states according to the {\it a priori} unknown joint posterior density $p(K,\vec{\theta}^{(k)} \{k = 1,\dots,K\}, \sigma_n|D)$.

\begin{figure}\label{f:plotl1}
\fbox{\resizebox{7.5cm}{!}{\includegraphics{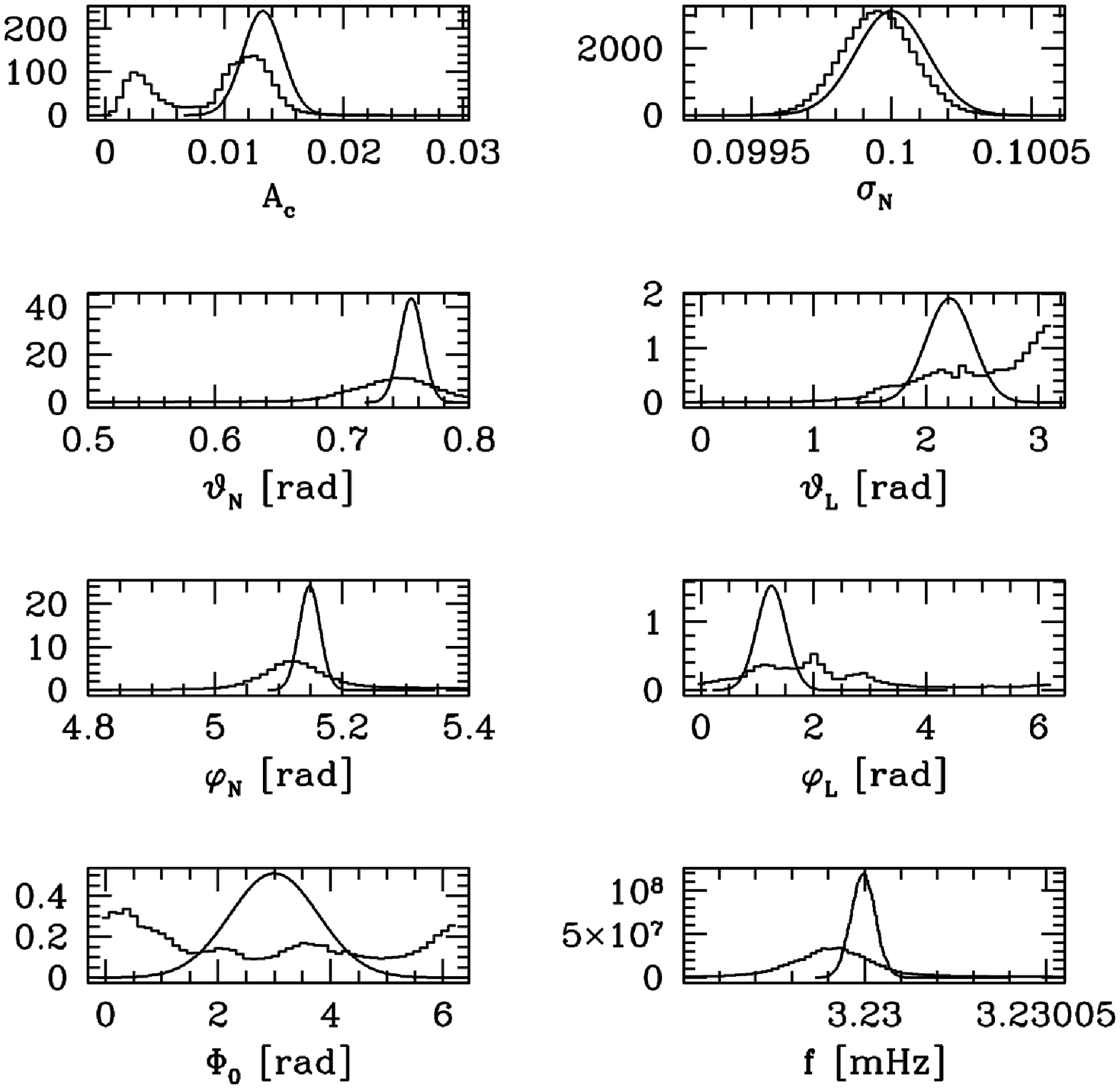}}}
\fbox{\resizebox{7.5cm}{!}{\includegraphics{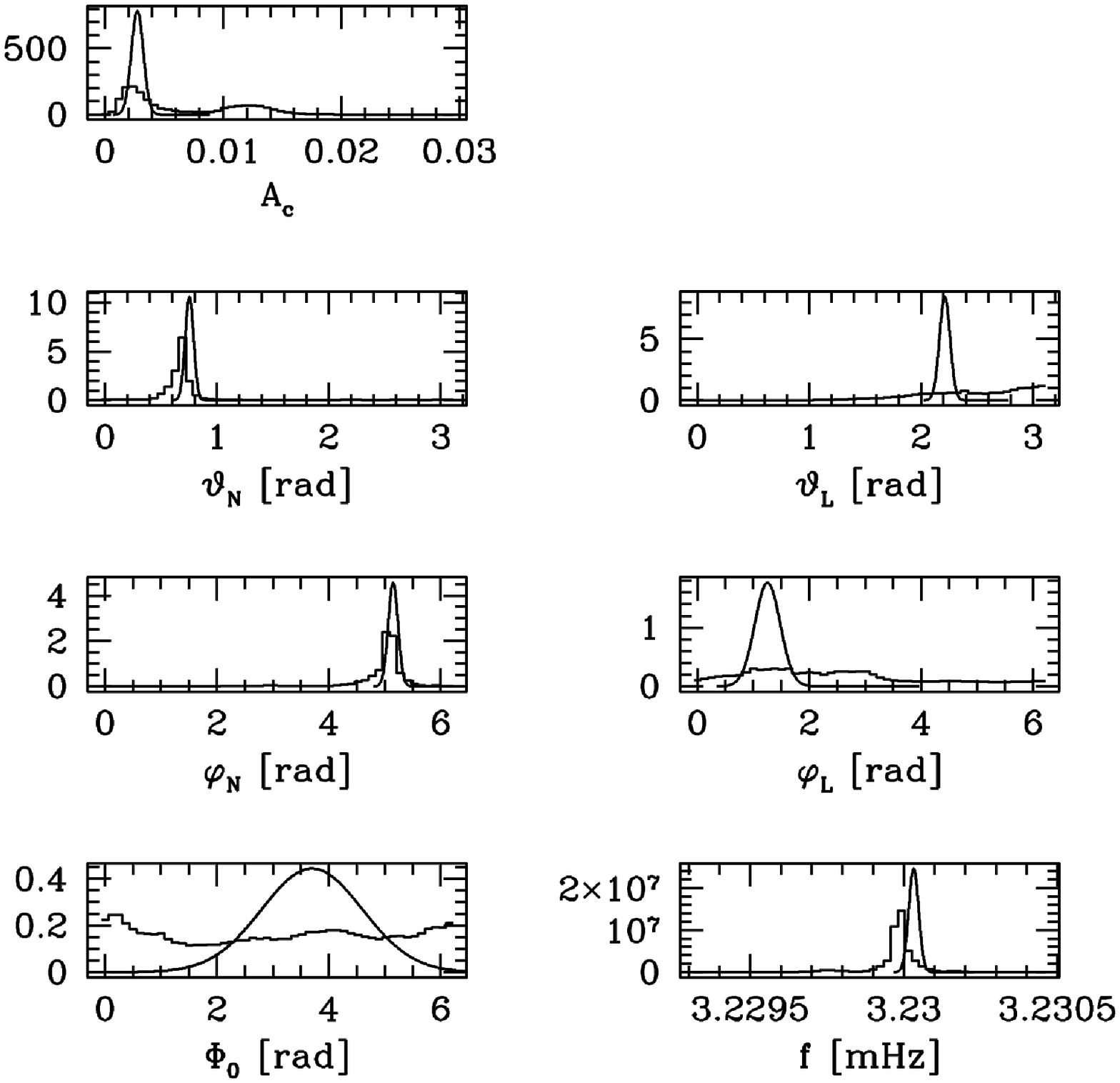}}}
\caption{Here we show the marginalised posterior distributions for the RJMCMC states in the simulation of two overlapping DWDs (black histogram). The plots on the left refer to the ES Cet-like source, and the plots on the right to the confusion source. Overlayed we show the theoretical Gaussian PDFs as computed from the variance-covariance matrix (grey continuous line) for a single source. See text for more details.}
\end{figure}

Our automatic RJMCMC sampler was able to disentangle the two signals in the data set with 68$\%$ probability; it estimated the presence of three signals and just one signal in the data stream with 32$\%$ and 0$\%$ probability, respectively. The marginalised distributions for $K=2$ are shown in Fig.~\ref{f:plotl1}, from which it is clear that we were also able to determine the correct noise level. The 14 parameters of the two signals were recovered in general to within one-to-two sampled standard deviations of the marginalised posterior PDFs, a value found by comparing the global mode of the distribution to the actual value of the parameter -- hereafter we will call this difference the ``bias'' or ``offset'' of the run. This offset was always present in our results and did not change significantly when we introduced, for example, longer chains and/or different initial starting values for the Markov chains. In order to obtain other quantitative indications for the performance of the sampler, we also computed the variance-covariance matrix for each source (under the assumption that the data stream contains a single signal). It is well known that the diagonal elements of the variance-covariance matrix provide a lower bound (the so-called Cramer-Rao bound) to the uncertainties characterising the parameter extraction; such bound becomes tight only in the case of high SNR, although the exact value at which this occurs depends on the signal properties \cite{Nich1998a}. Since we worked with overlapping sources with moderate SNR, the theoretical, single-source, high signal-to-noise ratio PDFs should be considered only as a broad guideline to evaluate the output from the actual analysis. In fact, Fig.~\ref{f:plotl1} shows some noticeable deviation from the theoretical predictions, such as the long tails in the PDFs for $\vartheta_N,\varphi_N$ and $f$, and the multi-modal/flat PDFs in $A$, $\Phi_0$, $\vartheta_L$ and $\varphi_L$. These features are due to the high degree of overlap of the two sources and strong correlations between these parameters.

\section{EMRI's and LISA data analysis strategies}
Extreme mass ratio inspirals (EMRIs) are one of the most exciting potential
sources of GWs for LISA. The source orbits are typically eccentric and inclined
to the orbital plane of the central black hole and the gravitational waveforms
they generate are richly coloured by precession of the perihelion, precession of
the orbital plane and inspiral. The complexity of the waveforms encodes a huge
amount of information about the source spacetime and orbit, but makes the
detection of these sources very challenging. A typical event has instantaneous
amplitude an order of magnitude below the noise in the LISA detector, but the 14
dimensional parameter space of possible signals is too large to permit detection
via fully coherent matched filtering \cite{gair04}. Two alternative detection
strategies have been examined to date with promising results --- a semi-coherent
matched filtering search \cite{gair04} and a time-frequency search algorithm
\cite{wengair05,gairwen05}. However, the former is very computationally
demanding and the latter does not provide much information about the source
parameters. Markov Chain Monte Carlo techniques are a quick and efficient way to
explore large dimensional parameter spaces, and so may well be a powerful tool
for EMRI detection. To investigate their effectiveness, we constructed a toy
model based on the prescription described by Barack and Cutler \cite{Barack2004b}, but using only the leading order terms for the evolution of the frequency, eccentricity etc. and a low eccentricity expansion of the source orbit. This model will be described in more detail elsewhere, but these simplifications allowed the waveform to be written explicitly as a function of time, which made it quicker to evaluate than the full model. For further simplicity, we fixed the three source parameters that specify the initial phases of the motion, i.e., we assumed the particle was always in the same place at the start of the observation. The remaining eleven parameters that described the EMRI system were the mass, $M$, and spin, $a$, of the central black hole, the mass, $\mu$, of the inspiralling object, the initial radial frequency, $\nu_0$, of the orbit, the inclination, $\iota$, of the orbit with respect to the orbital plane of the central black hole, the initial eccentricity, $e_0$, of the orbit, the distance, $d$, to the source, and the position, ($\vartheta_N,\varphi_N$), and orientation, ($\vartheta_L,\varphi_L$), of the massive black hole in the sky in ecliptic coordinates.

We used a time series data set containing one EMRI signal observed over a total observation time of 2$\times10^6$ seconds, sampled every 10 seconds. We implemented one Cutler Michelson observable (hI) in the long wavelength approximation \cite{Cutler1998a} to model the orbit and response of LISA. Noise was again taken to be stationary, white and Gaussian. To make the EMRI signal more realistic, we considered a typical EMRI signal from a source with $M=10^6\Ms$ and $\mu=10\Ms$, observed at a distance of 1 Gpc for one year. We then adjusted $\mu$ by a factor of the ratio of one year to our observation time ($3.1\times10^{7}/2\times10^{6}$), and increased the distance to 4 Gpc in order to match the SNR to that of the ``typical'' source ($\rho = 11.5$). Effectively, this procedure ``squeezes'' a typical one year EMRI event into the observation time that computation allowed. For this case study, we set up the RJMCMC assuming that the number of signals in the data set was known to be one, i.e., there was no search over the number of signals present in the data. We sampled $3\times 10^6$ Markov chain states.

\begin{figure}\label{f:plotl2}
\fbox{\resizebox{7.5cm}{!}{\includegraphics{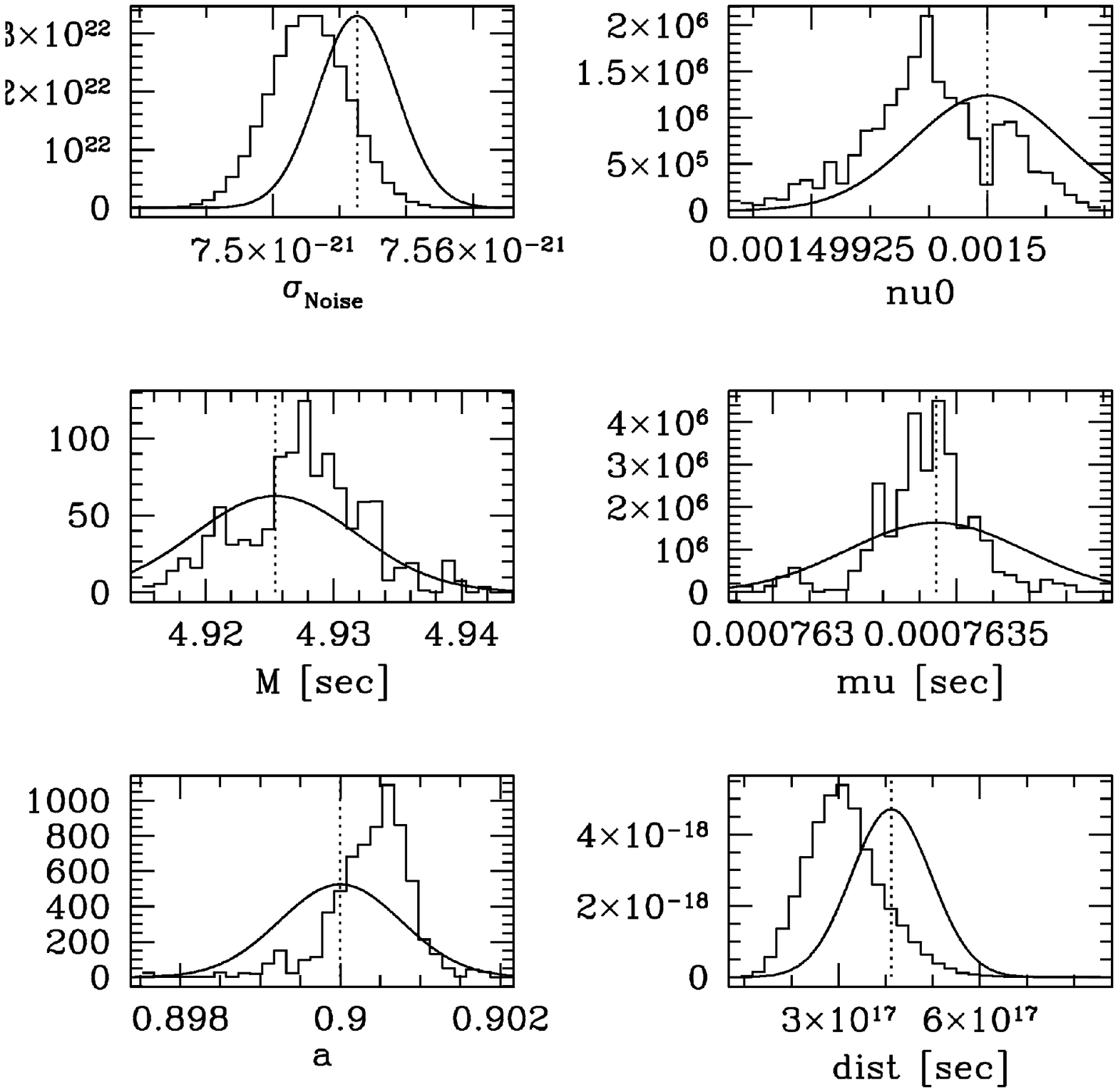}}}
\fbox{\resizebox{7.5cm}{!}{\includegraphics{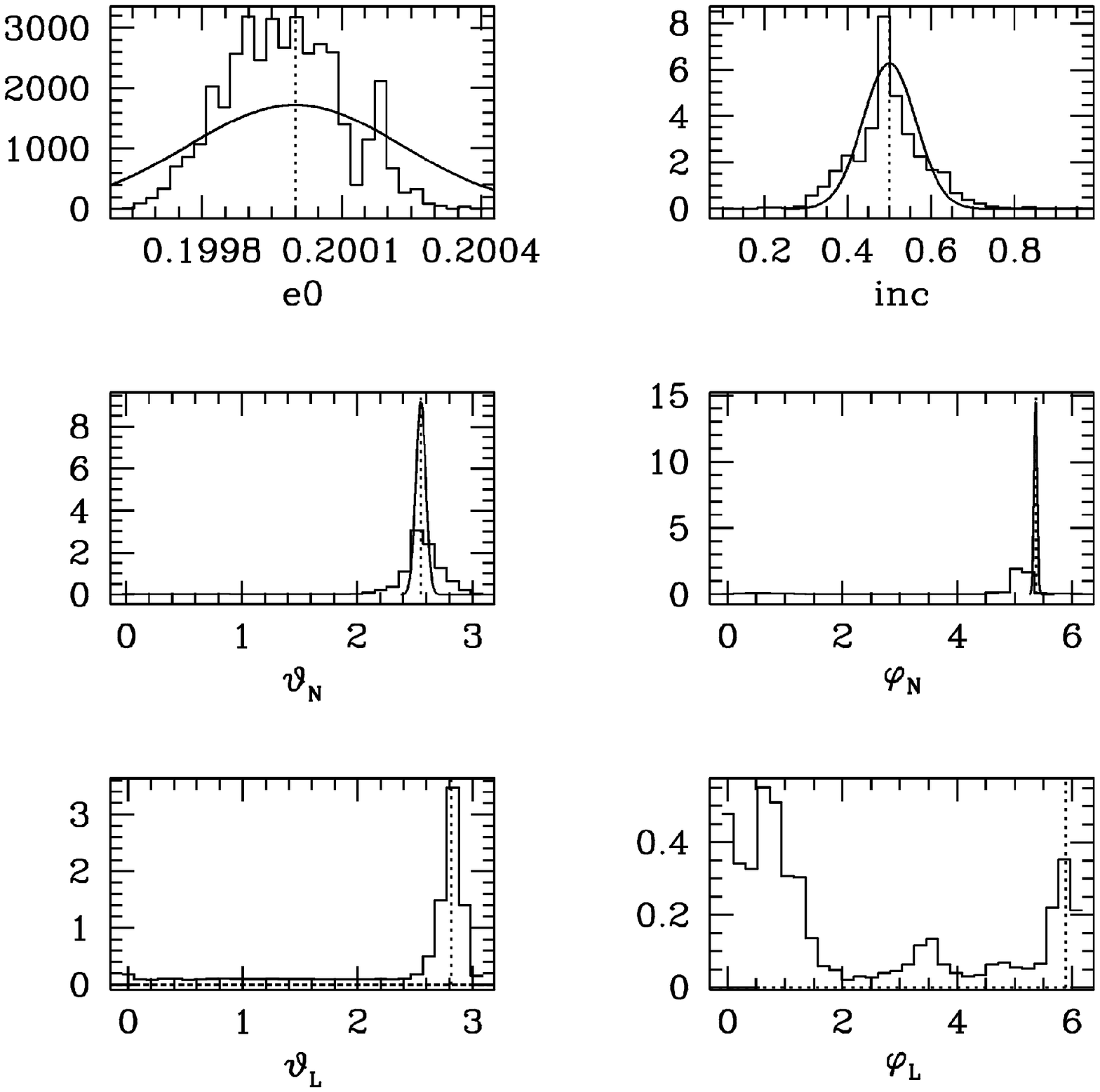}}}
\caption{Here we show the marginalised posterior distributions for our RJMCMC states in the simulation of EMRI signals (black histogram). Overlayed we show the theoretical lower bound on the uncertainties in parameter extractions derived from the (projected) Fisher information matrix (grey continuous Gaussian).}
\end{figure}

Using this setup, we were able to detect the signal in our data set without further problems. This is clear from the marginalised posterior distributions for the Markov chain states as shown in Fig.~\ref{f:plotl2}. The bias of the parameter extraction is clearly within one sampled standard deviation in all cases. In order to derive more information about the performance of the code, we again overlay lower bounds on the uncertainties for parameter extraction, determined from the single-source Fisher information matrix, with projection to account for marginalisation over correlated parameters (this is equivalent to the variance-covariance matrix). This calculation was difficult since, over the integration time used, the waveform parameters were highly correlated. This made the Fisher information matrix very degenerate and the subsequent projection/inversion subject to numerical errors. The shown predictions should therefore be considered just as guidelines. It is clear in Fig.~\ref{f:plotl2} that the marginalised posterior distributions generated by the RJMCMC are in general close to the predetermined Gaussian distributions for most of the parameters. However, in the case of $\nu_0$, $M$, $mu$, $a$, $d$, $\vartheta_L$ and $\varphi_L$ the recovered width of the Markov chain state distribution is narrow compared to the prediction. This might be explained by the uncertainties in our Fisher information matrix prediction. However, since the parameters are strongly correlated, we cannot exclude the possibility that these correlations prevented the chain from exploring the parameter space as freely as would be necessary to cover the edges of the joint posterior density. The steeper cut-off in the PDF of the eccentricity and the pedestal-like cut-offs in the PDFs for the masses might be indicative of this. In the future, we plan to examine the effectiveness of our algorithm more closely by using larger numbers of Markov chain states to ensure optimal coverage of the whole of the joint posterior density function. We have also carried out a similar simulation using the full Barack and Cutler model
\cite{Barack2004b} for the EMRI waveform, but over a shorter integration time ($5 \times 10^5$s). The results in that case were qualitatively similar, but the code takes approximately four times as long to complete the Markov Chain for a given observation time. This is because of the increased computational overhead associated with evaluating the full Barack and Cutler waveforms.

\section{Discussion}

We have introduced a new technique for computing the posterior probability density function of the source parameters in the context of LISA data analysis where the number of signals and the noise level in a data set are unknown. In particular, we have described the use of a new unsupervised and automatic Reversible Jump Markov Chain Monte Carlo technique.

We have examined two case studies to demonstrate the use of this sampler: highly overlapping signals from monochromatic white dwarf binary systems, and one complex signal from an EMRI. In the first study, we treated the number of sources as an unknown, but in the latter case we assumed that the number of sources was known to be one. In both cases we were able to successfully identify the signals in the data set automatically. In the case of the overlapping DWDs the recovered source parameters were within one or two sampled standard deviations of the true parameters. The marginalised posterior distributions deviate significantly from the single-source theoretical distribution predicted by the variance-covariance matrix, but this is to be expected due to the high degree of correlation in the system. A full quantitative characterisation of these features will be further investigated in the future. In the case of the EMRI, the sampler can identify and extract the values of the signal's source parameters automatically with high accuracy, with a bias always within one sampled standard deviation of the Markov chain state distribution. This is a promising result which suggests that MCMC may be a useful tool for the challenging task of EMRI detection. However, the waveform model used was too simplified and the observation time too short to allow us to make a generic statement at this stage.

It is interesting to consider why we see, but also expect, bias in our runs. Metropolis-Hastings sampling is an approximate method to evaluate the posterior PDF and as such always introduces bias. This bias can be divided into user-bias and systematical bias. Metropolis-Hastings creates a sample that approximates the real posterior density function asymptotically in time (equally to ongoing Markov chain state sampling). Since we end our program run after $10^7$ ($3 \times 10^6$) Markov chain states, we introduce bias since at this stage the method has only approximated the real posterior density to a limited degree. Although adaption as performed within AAP speeds up this convergence, this user-bias is still present. The systematical bias comes from two sources --- the noise in the data and, for our overlapping DWD problem, the high degree of overlap between the two binaries in the data. This systematic bias indicates how parameter extraction from a data set may never achieve ideal results, but will be subject to errors that depend on the signal to noise ratio and the complexity and make up of the data stream.

We plan to extend the current work in several ways. Although a parallel code was used to generate the results presented here, the numerical implementation is still too CPU intensive and not suitable for the actual analysis of a realistic LISA data set. In the future we plan to investigate and implement numerical optimisations of various aspects of the sampler to reduce the code run-times. For the EMRI study, such optimisation should allow us to analyse longer data streams, lasting 2 to 5 years. We plan to initially use the toy model to achieve these longer observation times, before switching to the more realistic Barack and Cutler model for further studies (and/or other waveforms that are currently becoming available). For the DWD study, the optimisation will allow us to explore the scenario in which many sources are present in the data set. We will also consider the full set of TDI channels~\cite{vallis05} and use simulated noise coloured according to the expected spectral density.

\begin{theacknowledgments}
AS wants to thank David Hastie, John Veitch and Graham Woan for helpful discussions. This work was supported in part by St.Catharine's College (JG).
\end{theacknowledgments}



\end{document}